\documentclass[aps,prc,twocolumn,superscriptaddress]{revtex4-2}
\usepackage{graphicx}
\usepackage{makecell}
\usepackage{color}
\usepackage[dvipsnames]{xcolor}
\usepackage[colorlinks=true,urlcolor=Blue,linkcolor=Blue]{hyperref}
\usepackage[all]{hypcap}
\usepackage{amsmath}
\usepackage[utf8]{inputenc}
\DeclareUnicodeCharacter{FF1A}{\text{:}}

\begin{document}


\title{Nuclear structure of dripline nuclei elucidated through precision mass measurements of $^{23}$Si, $^{26}$P, $^{27,28}$S, and $^{31}$Ar}

\author{Y.~Yu}\thanks{These authors contributed equally to this work.}
\affiliation{CAS Key Laboratory of High Precision Nuclear Spectroscopy, Institute of Modern Physics, Chinese Academy of Sciences, Lanzhou 730000, China}
\affiliation{School of Nuclear Science and Technology, University of Chinese Academy of Sciences, Beijing 100049, China}	

\author{Y.~M.~Xing}\thanks{These authors contributed equally to this work.}
\affiliation{CAS Key Laboratory of High Precision Nuclear Spectroscopy, Institute of Modern Physics, Chinese Academy of Sciences, Lanzhou 730000, China}
\affiliation{School of Nuclear Science and Technology, University of Chinese Academy of Sciences, Beijing 100049, China}

\author{Y.~H.~Zhang}
\email{Corresponding author: yhzhang@impcas.ac.cn}
\affiliation{CAS Key Laboratory of High Precision Nuclear Spectroscopy, Institute of Modern Physics, Chinese Academy of Sciences, Lanzhou 730000, China}
\affiliation{School of Nuclear Science and Technology, University of Chinese Academy of Sciences, Beijing 100049, China}	

\author{M.~Wang}
\email{Corresponding author: wangm@impcas.ac.cn}
\affiliation{CAS Key Laboratory of High Precision Nuclear Spectroscopy, Institute of Modern Physics, Chinese Academy of Sciences, Lanzhou 730000, China}
\affiliation{School of Nuclear Science and Technology, University of Chinese Academy of Sciences, Beijing 100049, China}

\author{X.~H.~Zhou}
\affiliation{CAS Key Laboratory of High Precision Nuclear Spectroscopy,   Institute of Modern Physics, Chinese Academy of Sciences, Lanzhou 730000, China}
\affiliation{School of Nuclear Science and Technology, University of Chinese Academy of Sciences, Beijing 100049, China}

\author{J.~G.~Li}
\affiliation{CAS Key Laboratory of High Precision Nuclear Spectroscopy,   Institute of Modern Physics, Chinese Academy of Sciences, Lanzhou 730000, China}
\affiliation{School of Nuclear Science and Technology, University of Chinese Academy of Sciences, Beijing 100049, China}

\author{H.~H.~Li}
\affiliation{CAS Key Laboratory of High Precision Nuclear Spectroscopy,   Institute of Modern Physics, Chinese Academy of Sciences, Lanzhou 730000, China}
\affiliation{School of Nuclear Science and Technology, University of Chinese Academy of Sciences, Beijing 100049, China}

\author{Q.~Yuan}
\affiliation{CAS Key Laboratory of High Precision Nuclear Spectroscopy, Institute of Modern Physics, Chinese Academy of Sciences, Lanzhou 730000, China}

\author{Y.~F.~Niu}
\affiliation{School of Nuclear Science and Technology, Lanzhou University, Lanzhou 730000, China}
\author{Y.~N.~Huang}
\affiliation{School of Nuclear Science and Technology, Lanzhou University, Lanzhou 730000, China}
\author{J.~Geng}
\affiliation{School of Nuclear Science and Technology, Lanzhou University, Lanzhou 730000, China}

\author{J.~Y.~Guo}
\affiliation{School of Physics and Optoelectronic Engineering, Anhui University, Hefei 230601, China}

\author{J.~W.~Chen}
\affiliation{State Key Laboratory of Nuclear Physics and Technology, School of Physics, Peking University, Beijing 100871, China}

\author{J.~C.~Pei}
\affiliation{State Key Laboratory of Nuclear Physics and Technology, School of Physics, Peking University, Beijing 100871, China}

\author{F.~R.~Xu}
\affiliation{State Key Laboratory of Nuclear Physics and Technology, School of Physics, Peking University, Beijing 100871, China}

\author{Yu.~A.~Litvinov}
\email{Corresponding author：y.litvinov@gsi.de}
\affiliation{GSI Helmholtzzentrum f{\"u}r Schwerionenforschung, Planckstra{\ss}e 1, 64291 Darmstadt, Germany}

\author{K.~Blaum}
\affiliation{Max-Planck-Institut f\"{u}r Kernphysik, Saupfercheckweg 1, 69117 Heidelberg, Germany}	

\author{G.~de Angelis}
\affiliation{INFN Laboratori Nazionali di Legnaro, viale dell’Università 2 Legnaro, Italy}	
\affiliation{CAS Key Laboratory of High Precision Nuclear Spectroscopy, Institute of Modern Physics, Chinese Academy of Sciences, Lanzhou 730000, China}

\author{I.~Tanihata}
\affiliation{School of Physics, Beihang University, Beijing 100191, China}
\affiliation{Research Center for Nuclear Physics (RCNP), Osaka University, Ibaraki Osaka 567-0047, Japan}

\author{T.~Yamaguchi}
\affiliation{Department of Physics, Saitama University, Saitama 338-8570, Japan}

\author{X.~Zhou}
\affiliation{CAS Key Laboratory of High Precision Nuclear Spectroscopy,  Institute of Modern Physics, Chinese Academy of Sciences, Lanzhou 730000, China}
\affiliation{School of Nuclear Science and Technology, University of Chinese Academy of Sciences, Beijing 100049, China}

\author{H.~S.~Xu}
\affiliation{CAS Key Laboratory of High Precision Nuclear Spectroscopy,   Institute of Modern Physics, Chinese Academy of Sciences, Lanzhou 730000, China}
\affiliation{School of Nuclear Science and Technology, University of Chinese Academy of Sciences, Beijing 100049, China}









\author{Z.~Y.~Chen}
\affiliation{CAS Key Laboratory of High Precision Nuclear Spectroscopy, Institute of Modern Physics, Chinese Academy of Sciences, Lanzhou 730000, China}
\affiliation{School of Nuclear Science and Technology, University of Chinese Academy of Sciences, Beijing 100049, China}	

\author{R.~J.~Chen}
\affiliation{CAS Key Laboratory of High Precision Nuclear Spectroscopy,   Institute of Modern Physics, Chinese Academy of Sciences, Lanzhou 730000, China}
\affiliation{GSI Helmholtzzentrum f{\"u}r Schwerionenforschung, Planckstra{\ss}e 1, 64291 Darmstadt, Germany}

\author{H.~Y.~Deng}
\affiliation{CAS Key Laboratory of High Precision Nuclear Spectroscopy,   Institute of Modern Physics, Chinese Academy of Sciences, Lanzhou 730000, China}
\affiliation{School of Nuclear Science and Technology, University of Chinese Academy of Sciences, Beijing 100049, China}

\author{C.~Y.~Fu}
\affiliation{CAS Key Laboratory of High Precision Nuclear Spectroscopy,   Institute of Modern Physics, Chinese Academy of Sciences, Lanzhou 730000, China}

\author{W.~W.~Ge}
\affiliation{CAS Key Laboratory of High Precision Nuclear Spectroscopy,   Institute of Modern Physics, Chinese Academy of Sciences, Lanzhou 730000, China}

\author{W.~J.~Huang}
\affiliation{Advanced Energy Science and Technology Guangdong Laboratory, Huizhou 516007, China}
\affiliation{CAS Key Laboratory of High Precision Nuclear Spectroscopy, Institute of Modern Physics, Chinese Academy of Sciences, Lanzhou 730000, China}

\author{H.~Y.~Jiao}
\affiliation{CAS Key Laboratory of High Precision Nuclear Spectroscopy,   Institute of Modern Physics, Chinese Academy of Sciences, Lanzhou 730000, China}
\affiliation{School of Nuclear Science and Technology, University of Chinese Academy of Sciences, Beijing 100049, China}

\author{Y.~F.~Luo}
\affiliation{CAS Key Laboratory of High Precision Nuclear Spectroscopy,   Institute of Modern Physics, Chinese Academy of Sciences, Lanzhou 730000, China}
\affiliation{School of Nuclear Science and Technology, University of Chinese Academy of Sciences, Beijing 100049, China}

\author{H.~F.~Li}
\affiliation{CAS Key Laboratory of High Precision Nuclear Spectroscopy,   Institute of Modern Physics, Chinese Academy of Sciences, Lanzhou 730000, China}

\author{T.~Liao}
\affiliation{CAS Key Laboratory of High Precision Nuclear Spectroscopy,   Institute of Modern Physics, Chinese Academy of Sciences, Lanzhou 730000, China}
\affiliation{School of Nuclear Science and Technology, University of Chinese Academy of Sciences, Beijing 100049, China}


\author{J.~Y.~Shi}
\affiliation{CAS Key Laboratory of High Precision Nuclear Spectroscopy,   Institute of Modern Physics, Chinese Academy of Sciences, Lanzhou 730000, China}
\affiliation{School of Nuclear Science and Technology, University of Chinese Academy of Sciences, Beijing 100049, China}

\author{M.~Si}
\affiliation{CAS Key Laboratory of High Precision Nuclear Spectroscopy,   Institute of Modern Physics, Chinese Academy of Sciences, Lanzhou 730000, China}

\author{M.~Z.~Sun}
\affiliation{CAS Key Laboratory of High Precision Nuclear Spectroscopy, Institute of Modern Physics, Chinese Academy of Sciences, Lanzhou 730000, China}

\author{P.~Shuai}
\affiliation{CAS Key Laboratory of High Precision Nuclear Spectroscopy, Institute of Modern Physics, Chinese Academy of Sciences, Lanzhou 730000, China}

\author{X.~L.~Tu}
\affiliation{CAS Key Laboratory of High Precision Nuclear Spectroscopy, Institute of Modern Physics, Chinese Academy of Sciences, Lanzhou 730000, China}


\author{Q.~Wang}
\affiliation{CAS Key Laboratory of High Precision Nuclear Spectroscopy,   Institute of Modern Physics, Chinese Academy of Sciences, Lanzhou 730000, China}

\author{X.~Xu}
\affiliation{CAS Key Laboratory of High Precision Nuclear Spectroscopy,   Institute of Modern Physics, Chinese Academy of Sciences, Lanzhou 730000, China}


\author{X.~L.~Yan}
\affiliation{CAS Key Laboratory of High Precision Nuclear Spectroscopy,   Institute of Modern Physics, Chinese Academy of Sciences, Lanzhou 730000, China}


\author{Y.~J.~Yuan}
\affiliation{CAS Key Laboratory of High Precision Nuclear Spectroscopy,   Institute of Modern Physics, Chinese Academy of Sciences, Lanzhou 730000, China}
\affiliation{School of Nuclear Science and Technology, University of Chinese Academy of Sciences, Beijing 100049, China}


\author{M.~Zhang}
\affiliation{CAS Key Laboratory of High Precision Nuclear Spectroscopy,  Institute of Modern Physics, Chinese Academy of Sciences, Lanzhou 730000, China}



\date{\today}



\begin{abstract}

Using the B$\rho$-defined isochronous mass spectrometry technique, we report the first determination of the $^{23}$Si, $^{26}$P, $^{27}$S, and $^{31}$Ar masses and improve the precision of the $^{28}$S mass by a factor of 11. Our measurements confirm that these isotopes are bound and fix the location of the proton dripline in P, S, and Ar. We find that the mirror energy differences of the mirror-nuclei pairs $^{26}$P-$^{26}$Na, $^{27}$P-$^{27}$Mg, $^{27}$S-$^{27}$Na, $^{28}$S-$^{28}$Mg, and $^{31}$Ar-$^{31}$Al deviate significantly from the values predicted assuming mirror symmetry. In addition, we observe similar anomalies in the excited states, but not in the ground states, of the mirror-nuclei pairs $^{22}$Al-$^{22}$F and $^{23}$Al-$^{23}$Ne. Using $ab~ initio$ VS-IMSRG and mean field calculations, we show that such a mirror-symmetry breaking phenomenon can be explained by the extended charge distributions of weakly-bound, proton-rich nuclei. When observed, this phenomenon serves as a unique signature that can be valuable for identifying proton-halo candidates.

\end{abstract}



\maketitle

A nucleus consists of $N$ neutrons and $Z$ protons confined in a finite size of several femtometers by nuclear forces.
Integral effects of strong and electromagnetic interactions determine the binding energy of the nucleus, which can be derived directly from the nuclear mass. 
Nuclear masses are crucial for revealing and explaining amazing nuclear structure phenomena\,\cite{Bohr-1998, YAMAGUCHI2021103882,BLAUM20061}, the appearance of new magic numbers~\cite{SORLIN2008602,PhysRevLett.84.5493,PhysRevLett.109.022501,PhysRevLett.114.202501} and disappearance of the conventional ones~\cite{SORLIN2008602,PhysRevLett.108.142501,MOTOBAYASHI19959,PhysRevLett.99.022503}, nucleon correlations\,\cite{Duguet-2001,Litvinov-2005a}, exotic decay modes\,\cite{BLANK2008403, Litvinov-2011,RevModPhys.84.567}, etc. 
Precision masses of proton-dripline nuclei can also be used to reveal  
mirror-symmetry breaking phenomena and the emergence of proton halos as will be demonstrated in this Letter.

Within the framework of isospin symmetry,
nuclear states are characterised by a total isospin $T$ and a projection $T_z=(N-Z)/2$. 
For an isospin multiplet, nuclei with exchanged numbers of neutrons and protons, i.e., the mirror nuclei, should have
an identical set of states if the nucleon-nucleon interaction were entirely charge symmetric.
Clearly, isospin is not a perfect symmetry: 
Protons and neutrons have different electric charges, their masses are slightly different (0.14\%) and their magnetic moments differ substantially in both magnitude and sign. 
The emergence of nuclear charge-symmetry breaking is therefore not at all surprising~\cite{Smirnova-2023}.
It is more the robust nature of isospin symmetry that is noteworthy, and those cases where deviations are found offer a chance to comprehend better the structure of nuclei. One of them is the Thomas-Ehrman shift (TES) which was initially observed in the {\it sd}-shell mirror nuclei~\cite{PhysRev.88.1109,PhysRev.81.412}.
In such nuclei the $l=0$ and $l=2$ states are nearly degenerate.
Since these orbits have different radial extensions, their Coulomb displacement energies (CDEs) are different.

Similarly, the formation of proton halos, characterised by the proton radial extensions, shall be reflected in the variation of CDEs.
In this context, $sd$-shell nuclei near the proton dripline are weakly bound 
and an extended spatial distribution of the valence protons is expected if the $\pi 2s_{1/2}$ single-particle orbit is dominantly occupied. 
Due to the loose binding and the absence of a centrifugal barrier for the proton $s_{1/2}$ shell, halo structures are more easily formed\,\cite{BROWN1996391,PhysRevC.53.R572,RevModPhys.76.215,TANIHATA2013215,TANIHATA2013215}. 
Indeed, enhanced reaction cross sections\,\cite{ZHANG2002303,PhysRevC.69.034326} and narrow momentum distributions\,\cite{PhysRevLett.81.5089} were reported as evidence for proton halos in $^{26,27,28}$P.
Further, large asymmetries in the Gamow-Teller transitions of $^{22}$Si/$^{22}$O\,\cite{PhysRevLett.125.192503} and $^{26}$P/$^{26}$Na\,\cite{PhysRevC.93.064320,sym13122278} mirror pairs were observed, providing arguments for the existence of proton halos in $^{22}$Al and $^{26}$P. Also the TESs were predicted to be larger for the ground states than for nearly all excited states in $^{26,27,28}$P\,\cite{PhysRevC.99.064330}. 

Prior to this work, there were no experimental mass values for $^{23}$Si, $^{26}$P, $^{27}$S, and $^{31}$Ar. 
This had hindered the identification of the proton dripline in these elements for which nuclear masses with an uncertainty of a few tens of keV are required. 
Furthermore, the knowledge of these masses is necessary for constraining astrophysical $(p,\gamma)$ reaction rates, e.g., $^{26}$P($p$,$\gamma$)$^{27}$S\,\cite{Hou-2023}, or for understanding the Gamow-Teller transition rates of 
$^{26}$P/$^{26}$Na mirror pairs\,\cite{PhysRevC.93.064320,sym13122278}. In this Letter, we report the precision mass measurements of $^{23}$Si, $^{26}$P, $^{27,28}$S, and $^{31}$Ar using the newly developed $B\rho$-defined isochronous mass spectrometry ($B\rho$-IMS)~\cite{PhysRevC.106.L051301,zhangm2023,Sun_2024}. Based on the new and available mass data and mirror symmetry in the $sd$-shell nuclei, we suggest a sensitive approach to identify proton-halo candidates.

The experiment was conducted at the Heavy Ion Research Facility in Lanzhou (HIRFL)~\cite{XIA200211,ZHAN2010694c}.
$^{36}$Ar$^{15+}$ ions were accelerated to an energy of $E/A\approx401$\,MeV by the main Cooler Storage Ring (CSRm).
The ion beam was fast-extracted and used to bombard a 15-mm thick beryllium target at the entrance of the fragment separator RIBLL2. 
Exotic nuclei were produced via fragmentation of $^{36}$Ar$^{15+}$. 
The reaction fragments emerging from the target were fully stripped of bound electrons, and thus had the charge $q=Qe=Ze$, with $e$ the elementary charge and $Q$ the charge number of the fragment. 
They were then in-flight separated from the primary beam with RIBLL2, 
and injected into the 128.8-m long experimental Cooler Storage Ring (CSRe). 

CSRe was tuned to the isochronous mode with $\gamma_t=1.339$, a machine parameter fixed by the ion-optics setting. 
The momentum acceptance of CSRe was $\pm 0.33 \%$. 
RIBLL2-CSRe was set to a central magnetic rigidity $B\rho=4.8417$\,Tm. 
In this setting, the isochronous condition was optimal for the nuclei with mass-to-charge ratio $m/Q\approx 1.73$ u (u is the atomic mass unit), providing the highest mass resolving power and transmission efficiency for $^{26}$P$^{15+}$.
Every 25\,s, a cocktail beam including the nuclides of interest was injected and stored in CSRe. 

Two identical time-of-flight (TOF) detectors were installed 18~m apart in one of the two straight sections of CSRe~\cite{ZHANG20141,XLYan2019}.
Each detector consists of a thin carbon foil ($\phi$40~mm, 18~$\rm\mu g/cm^2$) and a set of micro-channel plates (MCP). 
When an ion passed through the carbon foil, secondary electrons were released from the foil surface and 
guided to the MCP by electric and magnetic fields. 
Fast timing signals from the two MCPs were recorded by a digital oscilloscope at a sampling rate of 50~GHz. 

The measurement duration was 400~$\mu$s after an injection trigger, corresponding to $\sim 600$ revolutions of the ions in the ring.
From the timing signals, two sequences of time stamps for each stored ion were extracted\,\cite{XLTu2011N}. Then 
the revolution time and velocity were determined simultaneously using the procedure described in~\cite{XZhou2021v}. On average, about nine ions were stored simultaneouly in each injection. Particle identification was made following the procedures described in~\cite{XING2019}.

\begin{table*}[!htbp]
	\caption{Mass excesses obtained in this work ${\rm (ME_{IMS})}$. 
		Also listed are the number of events, the mass excesses in AME20 ${\rm (ME_{AME20})}$ with "\#" being the extrapolated ones~\cite{AME2020,Kondev_2021}, and the mass differences with respect to AME20 ($\Delta {\rm ME}$), to the mass predictions in~\cite{PhysRevC.105.034321} ($\Delta {\rm ME_{th1}}$) and to those obtained from improved Garvey-Kelson mass relations~\cite{PhysRevC.87.014313} ($\Delta {\rm ME_{th2}}$). The last two columns give, respectively, one- and two-proton separation energies ($S_p/S_{2p}$) and the average occupation number of valence protons in the $2s_{1/2}$ orbit, $N_p(2s_{1/2})$, obtained via the $ab~initio$ VS-IMSRG calculations~\cite{Li_2023,PhysRevC.107.014302}.}
	\centering
	\footnotesize
	\setlength{\tabcolsep}{4pt}
	\renewcommand{\arraystretch}{1.2}
	\begin{tabular}{lccccccccc}
		\hline
		Atom   &  events  &${\rm ME_{IMS}}$ & ${\rm ME_{AME20}}$ & $\Delta {\rm ME}$ & $\Delta {\rm ME_{th1}}$ & $\Delta {\rm ME_{th2}}$ & $S_p/S_{2p}$ &$N_p (2s_{1/2})$  \\
		&          & (keV)  & (keV) & (keV) & (keV) &  (keV) &(keV) & \\
		\hline
		$^{22}$Al$^{\rm a}$\footnotetext{The mass of $^{22}$Al has been used in \,\cite{Sun_2024} to test $ab~initio$ VS-IMSRG calculations} &  296  & $18103(10)$    & $18200(400)$\#      & $-97(400)   $  & 125(69)& 134(16)& 90(10)/3325(10)  & 0.246 \\
		$^{23}$Si  & 7               & $23537(119)$   & $23950(500)$\#    & $-413(514) $ & 297(145) & 243(155)&  1855(119)/1945(119) & 0.575 \\
		$^{26}$P   &  284            & $10998(11)$    & $10970(200)$\#    & $28(200)$  & $-129(69) $& $-205(12)$ & 118(15)/3531(11) &0.722 \\
		$^{27}$S   &  32            & $17418(39)$    & $17490(400)$\#    & $-72(400)$  & $-515(104)$ & $-516(41)$ & 869(41)/987(40) &1.421 \\
		$^{28}$S   &  189            & $4178(14)$    & $4070(160)$    & $108(160)$    & $-181(69)$ & $-209(15)$ & 2240(14)/3259(14) &1.184 \\
		$^{31}$Ar  &  188            & $11290(16)$  & $11330(200)$\#    & $-40(200)$   & $-277(80)$ & $-429(26)$ & 674(29)/194(21) & 1.760 \\
		
		\hline
	\end{tabular}
	\label{mass values table}
\end{table*}

Given the revolution time $t_{rev}$ and velocity $v$, the magnetic rigidity $B\rho$ and orbit length $C$ of the stored ions are determined according to 
\begin{equation}\label{eq:brho}
B\rho=\frac{m}{q}\gamma v,~~~~C=vt_{rev},
\end{equation}
where the Lorentz factor $\gamma=1/\sqrt{1-\beta^2}$ with $\beta$ being the velocity in units of the speed of light in vacuum. 
All particles with the same $B\rho$ must have the same mean $C$. 
Hence, the $B\rho(C)$ function characterizes the movement of all stored ions in the ring. 

\begin{figure}[htb]
	\includegraphics[angle=0,width=8.5 cm]{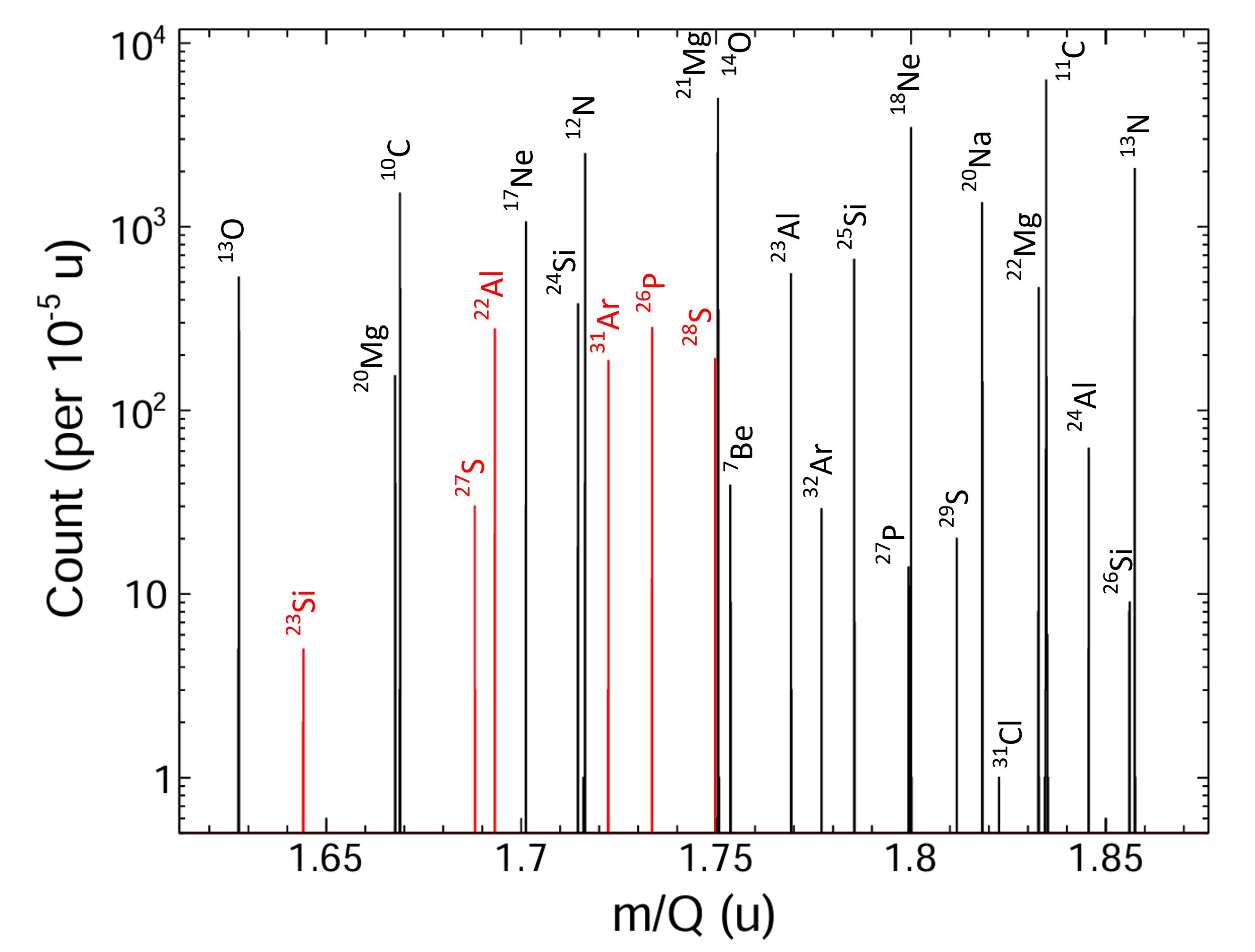}
	\caption{(Colour online) Part of the $m/Q$ spectrum zoomed in the region of $m/Q=$1.62 - 1.87\,u. Red peaks indicate the nuclei of present interest.   
		\label{mq-spectrum}}
\end{figure}

The experimental $\left\{B\rho_{\rm exp}^i,C_{\rm exp}^i,i=1,2, ...\right\}$ data were obtained according to Eq.~(\ref{eq:brho})
using the nuclei of well-known masses with uncertainties $\sigma <5$\,keV. 
The $B\rho (C)$ function was constructed by a least-squares fit.  
Then, the $m/q$ value of any stored ion, including the unknown-mass nuclei, can be directly obtained via 
\begin{equation}\label{eq:mvq}
\Big(\frac{m}{q}\Big)_{\rm exp}^i=\frac{B\rho (C_{\rm exp}^i)}{(\gamma v)_{\rm exp}^i}, ~~ i=1,~2,~3~...~.
\end{equation}
Equation~(\ref{eq:mvq}) is the basic formula of the $B\rho$-IMS, and the $B\rho (C)$ function is a universal calibration curve to be used for mass determination. The reader is referred to~\cite{zhangm2023} for more details.
All individual $m/Q$ values were put into a histogram forming an integrated $m/Q$ spectrum. Part of the spectrum is presented in Fig.~\ref{mq-spectrum}. 

\begin{figure}[t]
	\includegraphics[angle=0,width=8.5 cm]{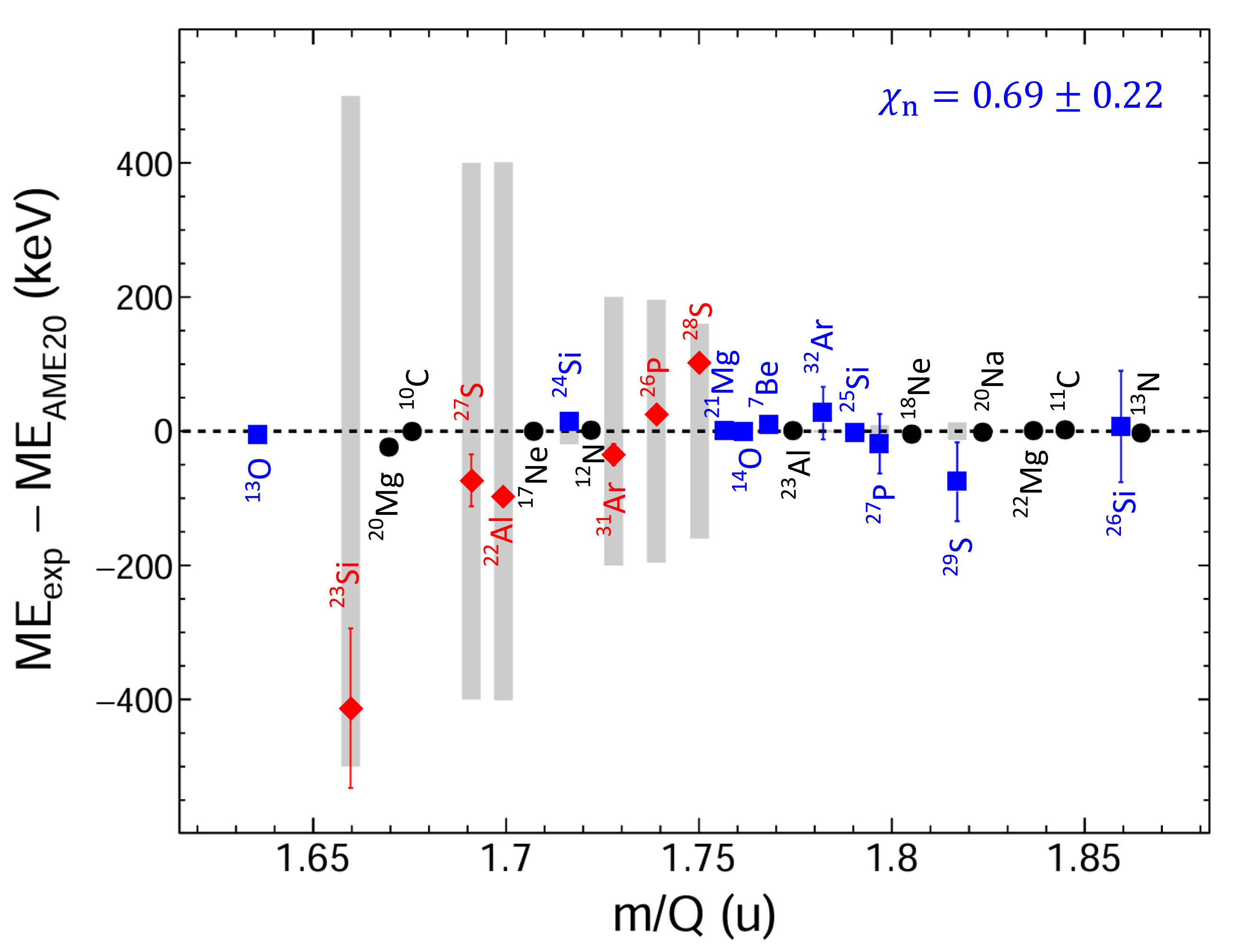}
	\caption{(Colour online) 
		Differences of mass excesses (MEs) obtained in this work and those from AME20~\cite{AME2020,Kondev_2021}. 
		The new MEs (red diamonds) are achieved by utilizing ten nuclides as calibrants (black circles). 
		The MEs of known-mass nuclides (blue squares) are re-determined in order to check the reliability of the analysis, which yield the normalized chi-square $\chi_{n}$ showing consistency with AME20. 
		The gray shades represent the $1\sigma$ uncertainties of AME20 values (not given if smaller than the size of the symbols).  
		\label{mass_result}}
\end{figure}

An individual $m/Q$ value and its uncertainty were obtained for each event. 
The weighted-average $m/Q$ was derived and converted~\cite{Zhang2018} into mass excess (ME), $ME=m-Au$.
In Table~\ref{mass values table} are listed the obtained results and their deviations from recent mass predictions~\cite{PhysRevC.87.014313,PhysRevC.105.034321} assuming mirror symmetry. The comparison with AME20~\cite{AME2020,Kondev_2021} is shown in Fig.~\ref{mass_result}. 

The masses of $^{23}$Si, $^{26}$P, $^{27}$S, and $^{31}$Ar are determined for the first time in this work, and the mass precision of $^{28}$S is improved by a factor of 11 compared to the literature value~\cite{AME2020,Kondev_2021}, 
reaching the relative precision of $\delta m/m\approx 5\times 10^{-7}$. 
We note that our mass of $^{22}$Al already reported in~\cite{Sun_2024} agrees, within $1\sigma$ standard deviation, with the more precise value of ME($^{22}$Al)=18092.5(3)\,keV measured using the LEBIT Penning trap mass spectrometer at FRIB~\cite{PhysRevLett.132.152501}.


The new mass values confirm that $^{22}$Al, $^{26}$P, $^{27}$S, and $^{31}$Ar are bound against one- and two-proton emission at a confidence level of more than $8\sigma $ (see Table~\ref{mass values table}). Since $^{21}$Al, $^{25}$P, $^{26}$S, and $^{30}$Ar are known to be particle unbound~\cite{PhysRevC.58.2831,PhysRevC.91.024307,doi:10.1142/S0218301311018216,PFUTZNER2023104050}, 
the location of the proton dripline is therefore fixed for these four elements. 

Mirror symmetry breaking effects are reflected in the nuclear binding energies, which can be translated into mirror energy differences (MEDs) defined as\,\cite{BROWN1996391,PhysRevC.97.034301}
\begin{equation}\label{eq:DSnp}
{\rm MED}_i(^A_ZX)=S_{in}(^A_{Z'}X')-S_{ip}(^A_{Z}X) ~~~(i=1,2), 
\end{equation}
where $S_{n}$ and $S_{2n}$ ($S_{p}$ and $S_{2p}$) represent one- and two-neutron (one- and two-proton) separation energies of the lower-$Z$, $^A_{Z'}X'$, (higher-$Z$, $^A_{Z}X$,) partner of the mirror pair, respectively. The extracted MEDs are assigned to the higher-$Z$ partners, $^A_{Z}X$.  
Given the nuclear charge symmetry and the long-range Coulomb force, 
such a quantity is a pure Coulomb effect and is sensitive to the proton spatial distribution which influences the variations of MEDs. 
It is expected that the larger spatial distribution in the proton-halo nucleus of the mirror pair makes it more bound, leading to a reduced 
MED.

\begin{figure}[htbp]
	\includegraphics[angle=0,width=8. cm]{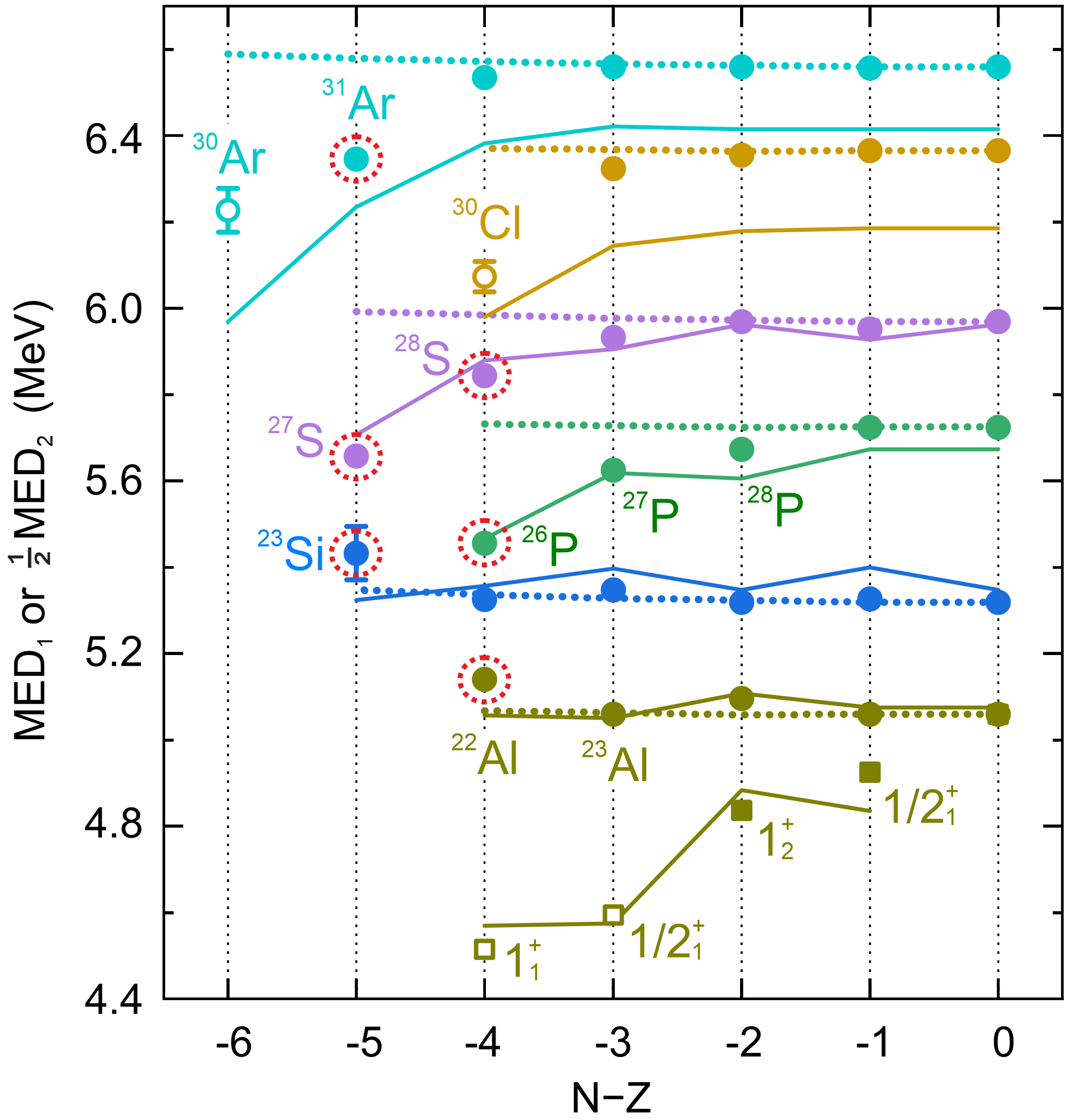}
	\caption{(Colour online) 
		Plot of mirror energy differences (MEDs) as a function of $N-Z$ of the proton-rich partners. MED$_2/2$ is used for plotting all MEDs in the same figure. 
		The dotted lines are calculations using Eqs.~(\ref{eq:Detlat_S}) and ({\ref{eq:Ec}}). 
		The thick solid lines are obtained from $ab~initio$ VS-IMSRG calculations. 
		The filled/open symbols indicate bound/unbound nuclei (or states). 
		Labels correspond to the proton-rich partners.		
		For $^{22}$Al, the data are shown for the ground and excited states.
		Our new values are marked by larger dotted circles.
		\label{delta_S}}
\end{figure}

The MEDs were extracted using the MEs in Table~\ref{mass values table}, the new mass of $^{27}$P~\cite{PhysRevC.108.065802}, and those in AME20~\cite{AME2020,Kondev_2021}. We used MED$_1$ when $Z$ is odd for the higher-$Z$ partner and MED$_2$ if $Z$ is even, such that the unbound nucleus is avoided in the extraction of MED. 
Figure~\ref{delta_S} presents the MEDs as a function of $N-Z$ of the higher-$Z$ partner, which is named in the figure to simplify the discussion hereafter. 
For completeness, MED$_2(^{30}{\rm Ar})$ and MED$_1(^{30}{\rm Cl})$ were derived using the recently reported $S_{2p}(^{30}{\rm Ar})$~\cite{PhysRevLett.115.202501,PhysRevC.97.034305} and $S_{p}(^{30}{\rm Cl})$~\cite{PhysRevC.105.044321}.

As expected from the mirror symmetry, Fig.~\ref{delta_S} shows constant MEDs for Al and Si isotopes.
However, there is a gradual decrease of MEDs for P, S, Cl, and Ar isotopes when moving towards the proton dripline, 
and the largest decreases are observed for $^{26}$P, $^{27}$S, and $^{31}$Ar. 
We emphasize that MEDs for the unbound nuclei $^{30}$Cl and $^{30}$Ar follow the decreasing trends as well.  

To understand these different systematic trends, we use the isobaric mass equation~\cite{PhysRev.147.735} 
\begin{equation}\label{eq:mass}
M(A,T,T_z)=M_0(A,T)+E_c(A,T,T_z)+T_z\Delta m_{nH},
\end{equation}
where $M_0(A, T)$ represents the charge-free nuclear mass, 
$E_c(A,T,T_z)$ the total charge-dependent energy of a nuclear state with isospin $T$ and its projection $T_z$, 
and $\Delta m_{nH}=782$ keV the mass difference between a neutron and $^1$H. 
From Eqs.~(\ref{eq:DSnp}) and (\ref{eq:mass}) it follows that
\begin{equation}\label{eq:Detlat_S}
\begin{aligned}
&{\rm MED}_i^{cal}=\Delta E_c(A,T)-\Delta E_c(A-i,T-i/2), \\
&\Delta E_c(A,T)=E_c(A,T,T_z=-T)-E_c(A,T,T_z=T), 
\end{aligned}
\end{equation}
with $E_c(A,T,T_z)$ given by~\cite{Nolen69} 
\begin{equation}\label{eq:Ec}
E_{c}=\{0.6Z^{2}-0.46Z^{4/3}-0.15[1-(-1)^{Z}]\}\times \frac{e^{2}}{r_{0}A^{1/3}},
\end{equation}
where $\Delta E_c$ is the Coulomb energy difference between the $T_z=T$ and $T_z=-T$ mirror nuclei.
Here we note that MED$_i^{cal}$ should exhibit the trends expected under the assumptions of mirror symmetry of the nuclear force and identical wave functions of the mirror nuclei. 

The calculated MEDs are plotted in Fig.~\ref{delta_S}. To overlay these lines onto the experimental values, $r_0$ was varied within $1.263\sim 1.292$ to match the $N=Z$ nuclide of each element. 
One sees that the experimental MEDs for Al and Si isotopes are well reproduced by the simple calculations. 
However, the gradual decrease of MEDs can not be reproduced for P, S, Cl, and Ar. 
The observed deviations from the general trends of MEDs, 
especially the prominent decreasing in $^{26}$P, $^{27}$S, and $^{31}$Ar, are obvious manifestation of mirror-symmetry breaking,
which is most probably induced by forming an exotic structure, a proton halo, in the weakly bound proton-rich partners. 

We take phosphorus isotopes as an example for the following discussion.
The proton halos in $^{26,27,28}$P~\cite{PhysRevLett.81.5089,ZHANG2002303,PhysRevC.69.034326,PhysRevC.93.064320,sym13122278} were attributed 
to the spatial extension of the $2s_{1/2}$ valence-proton wave function~\cite{BROWN1996391,PhysRevC.53.R572}. 
When approaching the proton dripline, the nucleus becomes less bound, and the $2s_{1/2}$-proton wave function is more extended than 
that of the deeply-bound $2s_{1/2}$ neutron in its neutron-rich partner.
This extra extension results in a reduction of the repulsive Coulomb energy of the nucleus and causes an energy shift (i.e., TES~\cite{PhysRev.88.1109,PhysRev.81.412,PhysRevC.99.064330}).
Hence, compared to the mirror-symmetry expectation, the $S_p$ is larger leading to the reduced MED. 
The gradual decrease of MEDs can therefore be regarded as a sensitive probe to identify 
proton-halo structure in nuclei. 
From Fig.~\ref{delta_S} the observed deviations provide a new indicator for proton-halo structures in $^{26,27}$P.

Given that the MEDs for S, Cl, and Ar have similar deviations as for P (see Fig.~\ref{delta_S}) 
and adopting the criteria mentioned above, we conclude that a similar structure change occurs in S, Cl, and Ar when approaching the proton dripline.
Thus, similar to $^{26}$P, the proton-halo structures are formed in $^{27,28}$S and $^{31}$Ar. 
Inquisitively, the extremely small value of $S_{2p}$($^{31}$Ar)$=194(21)$ keV suggests that $^{31}$Ar, similar to $^{17}$Ne~\cite{PhysRevLett.101.252502,PhysRevC.99.064308}, has a $2p$-proton halo with $^{29}{\rm S}+p+p$ Borromean structure,  
where none of the two-body subsystems is bound~\cite{RevModPhys.76.215,TANIHATA2013215}.

The proton-halo structures in weakly bound nuclei mentioned above are investigated using the state-of-the-art $ab~initio$ 
valence-space in-medium similarity renormalization group (VS-IMSRG) calculations
based on two- and three-nucleon interactions from chiral effective field theory~\cite{ PhysRevLett.118.032502,doi:10.1146/annurev-nucl-101917-021120}.
Details of the calculations were described in~\cite{Li_2023,PhysRevC.107.014302}.
The results show that not only the roughly constant MEDs for Al and Si isotopes but also the gradual decreasing trends for P, S, Cl, and Ar isotopes
are reproduced, see the solid lines in Fig.\,\ref{delta_S}.
This indicates that the nuclear structure and its variations when approaching the proton dripline are well captured by the theory. 
The extracted average proton occupation numbers in the $2s_{1/2}$ orbit, $N_p(2s_{1/2})$, are listed in Table~\ref{mass values table}. 
The small proton separation energies and dominant $\pi 2s_{1/2}$ occupations in $^{26,27}$P, $^{27,28}$S, and  $^{31}$Ar support the formation of proton-halo structures. 
 
\begin{figure}[t]
	\includegraphics[angle=0,width=8.8 cm]{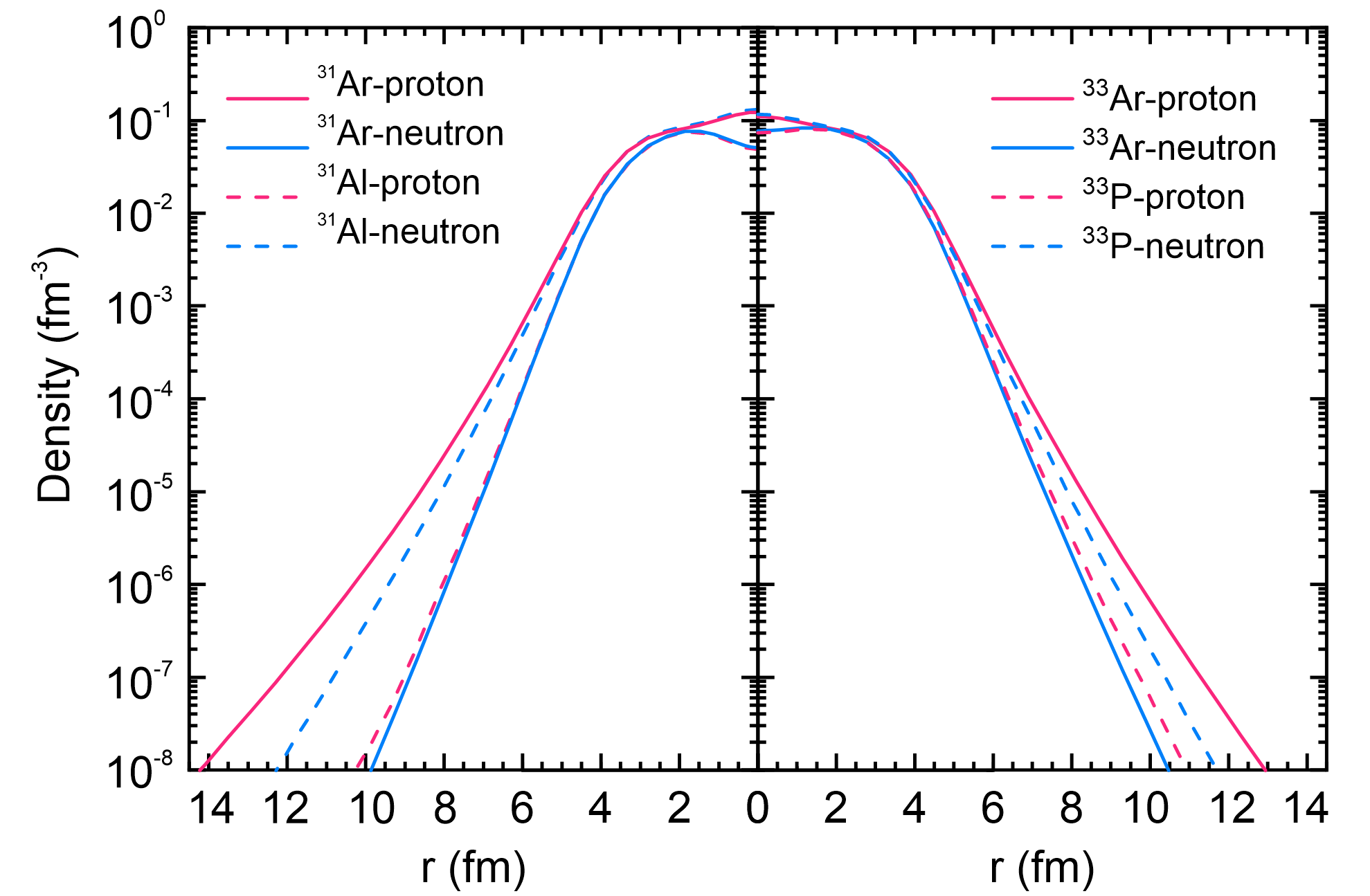}
	\caption{(Colour online) 
		Proton and neutron density distributions in proton-rich $^{31,33}$Ar and in the corresponding neutron-rich mirror partners $^{31}$Al and $^{33}$P. 
		\label{density2}}
\end{figure}

 The proton and neutron density distributions in the nuclei mentioned above 
 are calculated within the framework of self-consistent mean field theory using the deformed coordinate-space Hartree-Fock-Bogoliubov approach including continuum and axial deformations~\cite{PhysRevC.78.064306}.
 For P and S isotopes, similar features to those reported in~\cite{PhysRevC.53.R572,REN1999250,Liang_Yu_Jie_2009} are obtained. 
 We present in Fig.~\ref{density2} the density distributions for $^{31,33}$Ar along with their mirror partners $^{31}$Al and $^{33}$P.  One sees that the proton density distribution in $^{31}$Ar is more extended than the corresponding neutron density distribution in its mirror nucleus $^{31}$Al (see the left part of Fig.~\ref{density2}). Such a difference is less pronounced in the $^{33}$Ar/$^{33}$P mirror pair (see right part of Fig.~\ref{density2}). 
 The observed decreases of MEDs for $^{31}$Ar as well as for the particle-unbound $^{30}$Ar could be induced by the asymmetry of proton and neutron density distributions in the mirror pairs.
   
A peculiar case is $^{22}$Al. The $S_p(^{22}{\rm Al})$ is as low as 90(10) keV, but its MED follows the general trend predicted by Eq.~(\ref{eq:Detlat_S}), see Fig.~\ref{delta_S}. This can be understood by the small $\pi 2s_{1/2}$ occupation as presented in Table~\ref{mass values table}. 
Therefore, the ground state of $^{22}$Al is unlikely to be a proton-halo candidate. 
To further support this statement, 
we have extracted MEDs for the excited states involving the $\pi 1d_{5/2} \rightarrow \pi 2s_{1/2}$ excitation in $^{22,23,24,25}$Al, 
and the $\nu 1d_{5/2} \rightarrow \nu 2s_{1/2}$ excitation in the mirror partners of these Al isotopes. 
An obvious decreasing trend of MEDs is observed for the excited states, {see Fig.\,\ref{delta_S}.}
This indicates that the excited states, associated with predominant filling of the $\pi 2s_{1/2}$ orbit, form the proton-halo structures. 
This case is similar to $^{17}$F, in which the proton-halo structure is formed in the first $1/2^+$ excited state but not in the $5/2^+$ ground state\,\cite{PhysRevLett.79.3837,PhysRevLett.81.3341}. 

In conclusion, the nuclear masses of $^{23}$Si, $^{26}$P, $^{27,28}$S, and $^{31}$Ar were precisely measured using the newly developed $B\rho$-IMS
in CSRe. The high-precision results allowed us to determine for the first time the proton dripline for P, S, and Ar elements. 
We have investigated the MEDs in the $sd$-shell nuclei by taking the prediction of mirror-symmetry as a reference. It is found that MEDs are sensitive to structure changes 
and can be used as indicators of proton-halo structures in proton-rich nuclei. The observed significant mirror-symmetry breaking of MEDs in the $^{26}$P-$^{26}$Na, $^{27}$S-$^{27}$Na, $^{31}$Ar-$^{31}$Al mirror pairs 
is explained by proton halos in $^{26}$P, $^{27}$S, $^{31}$Ar. 
The same methodology is applied to excited states.
The analysis does not support the existence of proton halos in the ground states of $^{22-25}$Al, but instead, 
halos could be formed in their excited states involving $\pi 1d_{5/2}\rightarrow \pi 2s_{1/2}$ excitation.
Although our interpretation of the experimental data is supported by dedicated theoretical calculations and is overall consistent, precision nuclear charge radii measurements as e.g. by collinear laser spectroscopy~\cite{PhysRevLett.101.252502} would be important to reinforce the present interpretation.

\begin{acknowledgments}
	
	The authors thank the staff of the accelerator division of IMP for providing stable beam. 
	Fruitful discussions with Karlheinz Langanke, Ivan Mukha, Thomas Neff, Wilfried N{\"o}rtersh{\"a}user, and Achim Schwenk are greatly acknowledged.
	This work is supported in part by the Strategic Priority Research Program of Chinese Academy of Sciences (Grant No. XDB34000000), the National Key R\&D Program of China (Grant Nos. 2023YFA1606401, 2021YFA1601500), the NSFC (Grants No. 12135017, No. 12121005, No. 12335007, No. 11961141004, No. 12035001, No. 12305126,  No. 12205340, No. 12175281, No. 12347106), the Youth Innovation Promotion Association of Chinese Academy of Sciences
	(Grants No. 2021419 and No. 2022423), and the Gansu Natural Science Foundation (Grant No. 22JR5RA123, No. 23JRRA614). The theoretical \textit{ab initio} calculations in this work were done on Hefei advanced computing center. 
	
\end{acknowledgments}



%

\end{document}